\documentclass[preprint,prl,showpacs,groupedaddress,10pt,twocolumn]{revtex4}
\usepackage{amssymb}
\usepackage{graphics}
\usepackage{psfig}

\begin{document}

\preprint{}
\title[]{Rapid optimization of working parameters of microwave-driven
multi-level qubits for minimal gate leakage}
\author{Zhongyuan Zhou$^{1,2}$}
\author{Shih-I Chu$^{1}$}
\author{Siyuan Han$^{2}$}
\affiliation{$^{1}$Department of Chemistry, University of Kansas, Lawrence, KS 66045\\
$^{2}$Department of Physics and Astronomy, University of Kansas, Lawrence,
KS 66045}

\begin{abstract}
We propose an effective method to optimize the working parameters (WPs) of
microwave-driven quantum logical gates implemented with multi-level physical
qubits. We show that by treating transitions between each pair of levels
independently, intrinsic gate errors due primarily to population leakage to
undesired states can be estimated accurately from spectroscopic properties
of the qubits and minimized by choosing appropriate WPs. The validity and
efficiency of the approach are demonstrated by applying it to optimize the
WPs of two coupled rf SQUID flux qubits for controlled-NOT (CNOT) operation.
The result of this independent transition approximation (ITA) is in good
agreement with that of dynamic method (DM). \ Furthermore, the ratio of the
speed of ITA to that of DM scales exponentially as $2^{n}$ when the number
of qubits $n$ increases.
\end{abstract}

\received[Received: ]{June 29, 2005}
\maketitle

A practical quantum computer would be comprised of a large number of coupled
qubits and the coupled qubits must be kept in high-degree quantum coherence
states for sufficiently long time. During the past decade, significant
progress has been made on physical implementation of quantum computation.
High-degree quantum coherence has been demonstrated experimentally in
systems such as trapped ions \cite{Cirac95,Monroe96}, nuclear spins \cite%
{Gershenfeld97,Jones98}, atoms in optical resonators \cite{Turchette95}, and
photons in microwave cavities \cite{Brune96}. However, it seems quite
difficult to realize a large number of coupled qubits using these systems.
Meanwhile, solid-state qubits are of particular interest because of their
advantages of large-scale integration, flexibility in design, and easy
connection to conventional electronic circuits \cite{Mooij1999}. Of those,
qubits based on superconducting devices have recently attracted much
attention \cite{Makhlin2001} as manipulation of quantum coherent states
being successfully demonstrated in a variety of single qubits \cite%
{Nakamura1999,Vion2002,Yu2002,Martinis2002,Chiorescu2003} and coupled
two-qubit systems \cite%
{Pashkin2003,Yamamoto2003,Berkley2003,Mooij2004,Mooij2005}.

However, the solid-state qubits demonstrated in experiments so far all have
relatively short coherence time and high probability of gate errors \cite%
{Nakamura1999,Vion2002,Yu2002,Martinis2002,Chiorescu2003,Pashkin2003,Yamamoto2003,Mooij2004}%
. One of the causes of these problems is extrinsic gate error arising from
interaction between the environment and qubits resulting in decoherence,
such as dephasing and relaxation \cite{Makhlin2001}. Another cause is
intrinsic gate error resulting from population leakage to undesired states
due to the typical multi-level structures of solid-state qubits \cite%
{Fazio99,Zhou2002}. The intrinsic gate error is crucial since it not only
contributes to additional decoherence but also determines the ultimate
performance of the quantum gates and cannot be eliminated by reducing the
environment caused decoherence.

The leakage of a gate can be characterized quantitatively by summing up the
maximum transition probabilities to all undesired states of a multi-level
qubit \cite{Fazio99,Zhou2002}. It depends strongly on energy level structure
and transition matrix elements, i.e., spectroscopic properties of the
multi-level qubit determined completely by device parameters (DPs) and
external control parameters of the qubit which we call working parameters
(WPs) for simplicity. For instance, inductance (capacitance) of the
superconducting flux (charge) qubit is a DP while external flux (gate
voltage) is a WP. For the multi-level qubit with given DPs, the gate leakage
is sensitive to their WPs and thus can be minimized by the use of
appropriate WPs.

Conventionally, the leakage is calculated from transition probabilities to
all undesired states by numerically solving the time-dependent Schr\"{o}%
dinger equation (TDSE) \cite{Zhou2002, Zhou2004}. However, this kind of
dynamic method (DM) not only needs complicated numerical algorithms but also
a large amount of computational resources. For instance, optimizing the WPs
of a many-qubit network needed for a practical quantum computer using DM may
only be possible with \textit{ad hoc} powerful quantum computers in the
future. Thus a much faster approach is highly desirable.

In this Letter, we propose a very fast method to minimize the leakage of
microwave-driven quantum logical gates by choosing appropriate WPs in a
system of multi-level qubits with their DPs given \textit{in prior}. We
consider the case of weak microwave fields only since strong fields usually
cause many additional types of intrinsic gate errors and thus should be
avoided in general \cite{Zhou2002}. Our method is based on an independent
transition approximation (ITA) in which transitions in multi-level qubits
are treated independently. The leakage is estimated using the spectroscopic
properties obtained by solving the eigenvalue equation of and minimized by
optimizing the WPs of the qubits. The method is applied to minimize leakage
of controlled-NOT (CNOT) gate implemented with coupled rf superconducting
quantum interference device (SQUID) flux qubits. The result is in good
agreement with that obtained from DM. More importantly, the ITA is scalable
as the number of qubit $n$ increases because the ratio of the speed of ITA
to that of DM scales exponentially as $2^{n}$.

A microwave-driven gate is realized via coherent transitions between the
computational states of qubits interacting with microwave fields. In
general, correlation and interference between transitions cannot be ignored
and transition probability can only be computed accurately by solving TDSE.
However, in weak fields, which is the case considered here, only the
transitions between levels with which the microwave field is resonant or
nearly resonant are significant. For a given level, if the level spacings
between it and all other levels are sufficiently different, the correlation
and interference between the transitions from this level have negligible
effect. Hence, each transition is expected to take place independently and
the two levels involved can thus be treated as if they are isolated from the
others. For each of the two-level sub-systems interacting with a rectangular
pulse, the transition probability can be approximated by an analytical
expression using the rotating-wave approximation (RWA). Assuming the
rectangular pulse is $\mathbf{\epsilon }\left( t\right) =\mathbf{\epsilon }%
_{0}\cos \left( \omega t\right) $, the interaction between the qubit and the
pulse is then $V(t)=-\mathbf{\mu \text{\textperiodcentered }\epsilon }\left(
t\right) $, where $\mathbf{\epsilon }_{0}$ and $\omega $ are the amplitude
and frequency of the pulse and $\mathbf{\mu }$ is the dipole moment operator
of the qubit. The maximum transition probability from state $i$ to $j$ in an 
$N$-photon process, $\mathcal{P}_{ij}$, is \cite{Kmetic86}%
\begin{equation}
\mathcal{P}_{ij}=\Omega _{ij}^{2}/\left( \mathcal{D}_{ij}^{2}+\Omega
_{ij}^{2}\right) ,  \label{n3}
\end{equation}%
where, $\Omega _{ij}=2\mathbf{\mu }_{ij}$\textperiodcentered $\mathbf{%
\epsilon }_{0}NJ_{N}(y_{ij})/y_{ij}$ is the Rabi frequency for the $N$%
-photon resonance, $\mathcal{D}_{ij}=\Delta E_{ij}/\hbar -N\omega $\ is the
detuning, $\mathbf{\mu }_{ij}=\left\langle i\left\vert \mathbf{\mu }%
\right\vert j\right\rangle $ is the dipole transition matrix element, $%
\Delta E_{ij}=\left\vert E_{j}-E_{i}\right\vert $ is the level spacing
between the states $i$ and $j$, $J_{N}$ is the Bessel function of integer
order $N$, and $y_{ij}=\mathbf{d}_{ij}$\textperiodcentered $\mathbf{\epsilon 
}_{0}/\hbar \omega $ with $\mathbf{d}_{ij}=\mathbf{\mu }_{jj}-\mathbf{\mu }%
_{ii}$. Note that $\mathcal{P}_{ij}$, $\Omega _{ij}$, and $\mathcal{D}_{ij}$
depend on the number of photons $N$. In weak fields characterized by $\Omega
_{ij}\ll \omega $ only transitions with small $N$ are important. Since the
leakage is calculated from the maximum transition probabilities which are
independent of pulse shape in weak fields \cite{Shore-1} the optimized WPs
obtained using rectangular pulse are also valid for other pulse shapes.

For the microwave-driven gate considered here, the microwave is resonant
with a pair of computational states. According to Eq. (\ref{n3}), the
leakage is suppressed if $\mathcal{D}_{ij}\gg \Omega _{ij}$ holds for all
unintended transitions. This condition may be satisfied by using
sufficiently weak fields and/or qubits with proper spectroscopic properties.
However, the use of exceedingly weak fields will make the gate very slow.
Even so the unintended transitions will still cause large leakage at or near
resonance where $\mathcal{D}_{ij}\simeq 0$ and $\mathcal{P}_{ij}$ $\simeq 1$
regardless of intensity of the fields. Thus the leakage can be greatly
reduced by setting the level spacings of the qubits substantially detuned
from the microwave frequency and/or the dipole transition matrix elements
sufficiently small for all unintended transitions.

To investigate the leakage in a multi-level qubit, we study what happens
when the microwave acts on each of the computational states, which we call a
component of the gate. For an $n$-bit gate, there are $M=2^{n}$ components.
The leakage of the $i$th component is defined as $\eta _{i}=\sum_{k}P_{ik}$,
where $P_{ik}$ is the maximum probability of occupying an undesired state $k$
through all possible multi-photon transitions and the sum is over all the
undesired states including all non-computational states as well as those
computational states to which no transition is intended. The leakage of the $%
n$-bit gate, $\eta $, is then defined as $\eta =\max (\eta _{1},\eta
_{2},\cdots \eta _{M})$ which is a function of the $\nu $ WPs $q_{1}$, $%
q_{2} $, $\cdots $, and $q_{\nu }$. The set of WPs, $q_{1}^{\text{opt}}$, $%
q_{2}^{\text{opt}}$, $\cdots $, and $q_{\nu }^{\text{opt}}$ is obtained by
minimizing the leakage of gate, namely, $\eta \left( q_{1}^{\text{opt}%
},q_{2}^{\text{opt}},\cdots ,q_{\nu }^{\text{opt}}\right) \equiv \min \left[
\eta \left( q_{1},q_{2},\cdots ,q_{\nu }\right) \right] $.

As an example, we apply ITA to optimize the WPs of coupled rf SQUID flux
qubits for CNOT gate. Each rf SQUID consists of a superconducting loop of
inductance $L$\ interrupted by a Josephson tunnel junction characterized by
its critical current $I_{c}$ and shunt capacitance $C$ \cite{Danilov1983}. A
flux-biased rf SQUID with total magnetic flux $\Phi $ enclosed in the loop
is analogous to a "flux" particle of mass $m=C\Phi _{0}^{2}$, where $\Phi
_{0}$ $=h/2e$ is the flux quantum. The Hamiltonian is $h\left( x\right)
=p^{2}/2m+m\omega _{LC}^{2}\left( x-x_{e}\right) ^{2}/2-E_{J}\cos \left(
2\pi x\right) $. Here, $x=\Phi /\Phi _{0}$ is the canonical coordinate of
the \textquotedblleft flux\textquotedblright\ particle, $p=-i\hbar \partial
/\partial x$ is the canonical momentum conjugate to $x$, $E_{J}=\hbar
I_{c}/2e=m\omega _{LC}^{2}\beta _{L}/4\pi ^{2}$ is the Josephson coupling
energy, $\beta _{L}=2\pi LI_{c}/\Phi _{0}$ is the potential shape parameter, 
$\omega _{LC}=1/\sqrt{LC}$ is the characteristic frequency of the SQUID, and 
$x_{e}=\Phi _{e}/\Phi _{0}$ is the normalized external flux.

The coupled rf SQUID qubits comprise two rf SQUID qubits: a control qubit
and a target qubit coupled via their mutual inductance $M$ \cite%
{Mooij1999,zhou-asc2004}. For simplicity, we assume that the two SQUIDs are
identical: $C_{i}=C$, $L_{i}=L$, and $I_{ci}=I_{c}$ for $i=1$ and $2$. The
Hamiltonian of the coupled SQUID qubits is $H\left( x_{1},x_{2}\right)
=h\left( x_{1}\right) +h\left( x_{2}\right) +h_{12}\left( x_{1},x_{2}\right) 
$, where $x_{i}$, $x_{ei}$, and $h\left( x_{i}\right) $ ($i=1$ and $2$) are
the canonical coordinate, normalized external flux, and Hamiltonian of the $%
i $th single SQUID qubit and $h_{12}$ is the interaction between the qubits
given by $h_{12}(x_{1},x_{2})=m\omega _{LC}^{2}\kappa \left(
x_{1}-x_{e1}\right) \left( x_{2}-x_{e2}\right) $. Here, $\kappa =M/L$ is the
coupling constant. Note that each SQUID qubit is a multi-level system. The
eigenstate $|n)$ and eigenenergy $E_{n}$ of the coupled qubits are computed
by numerically solving the eigenvalue equation of $H(x_{1},x_{2})$ using the
two-dimensional Fourier-grid Hamiltonian method \cite{Chu1990}. They are
functions of the WPs $x_{e1}$, $x_{e2}$, and $\kappa $ for given device
parameters $L$, $C$, and $I_{c}$ \cite{Filippov2003,Brink05}. For weak
coupling $\kappa \ll 1$, the eigenstate of the coupled qubits, denoted by $%
|n)=\left\vert ij\right\rangle $, can be well approximated by the product of
the control qubit's state $\left\vert i\right\rangle $ and the target
qubit's state $\left\vert j\right\rangle $, $|n)=\left\vert i\right\rangle
\left\vert j\right\rangle $. When biased at $x_{e1},$ $x_{e2}\approx 1/2$
the potential of coupled SQUID qubits have four wells \cite%
{Mooij1999,zhou-asc2004}. The lowest eigenstates in each of the four wells,
denoted as $|1)=\left\vert 00\right\rangle $, $|2)=\left\vert
01\right\rangle $, $|3)=\left\vert 10\right\rangle $, and $|4)=\left\vert
11\right\rangle $, are used as the computational states of the coupled SQUID
qubits. For SQUIDs with $L=100$ pH, $C=40$ fF, and $\beta _{L}=1.2$, the
energy levels and level spacings versus $x_{e2}$ and $\kappa $ are plotted
in \textrm{Fig. 1(a)}$\mathrm{-}$\textrm{(d)}, respectively. It is shown
that both the energy levels and level spacings are sensitive to the WPs $%
x_{e2}$ and $\kappa $. When $x_{e2}<0.49877$ or $x_{e2}>0.50123$ in \textrm{%
Fig. 1(a)} or $\kappa >7.5\times 10^{-3}$ in \textrm{Fig. 1(c) }the energy
level structures become quite complicated. \textrm{Fig. 1(b) and Fig. 1(d)}
also show that in certain regions of the WP space the level spacings become
crowded, meaning that they are degenerate or nearly degenerate, which may
result in significant intrinsic gate errors.

\begin{figure}[ptb]
\begin{center}
\psfig{file=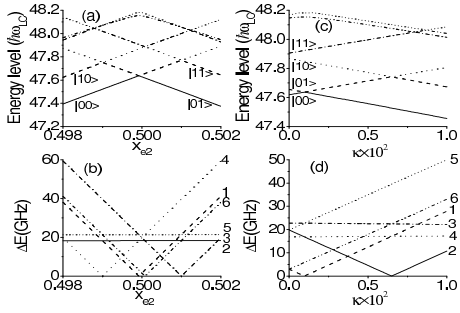,height=4cm,width=6cm}
\end{center}
\caption{Energy levels and level spacings of the coupled rf SQUID flux
qubits. (a) the energy levels and (b) the level spacings vs. $x_{e2}$ for $%
x_{e1}=0.499$ and $\protect\kappa =5\times 10^{-4}$, (c) the energy levels
and (d) the level spacings vs. $\protect\kappa $ for $x_{e1}=0.499$ and $%
x_{e2}=0.49985$. The lines 1 to 6 in (b) and (d) represent the level
spacings $\Delta E_{12}$, $\Delta E_{13}$, $\Delta E_{14}$, $\Delta E_{23}$, 
$\Delta E_{24}$, and $\Delta E_{34}$, respectively.}
\end{figure}

Controlled two-bit gates can be realized by applying a resonant microwave
pulse to the target qubit. The interaction between the microwave and the
coupled qubits can be written as $V\left( x_{1},x_{2},t\right) =m\omega
_{LC}^{2}\left[ \left( x_{2}-x_{e2}\right) +\kappa \left(
x_{1}-x_{e1}\right) +x_{m}/2\right] x_{m}$, where $x_{m}\left( t\right)
=x_{m0}\cos \left( \omega t\right) $ is the magnetic flux (normalized to $%
\Phi _{0}$) coupled to the target qubit from the microwave with amplitude $%
x_{m0}$ and frequency $\omega $. For the CNOT gate, a $\pi $-pulse with $%
\omega =\Delta E_{34}/\hbar $, where $\Delta E_{34}=\left\vert
E_{4}-E_{3}\right\vert $, is used. Populations of the states $\left\vert
10\right\rangle $ and $\left\vert 11\right\rangle $ are exchanged after the $%
\pi $-pulse only if the initial state is $\left\vert 10\right\rangle $ or $%
\left\vert 11\right\rangle $ or a linear combination of them. Note that in
this case in addition to the non-computational states, the undesired states
also include the computational states $\left\vert 00\right\rangle $ and $%
\left\vert 01\right\rangle $. Using Eq. (\ref{n3}) and following the steps
described above, the leakage of the CNOT gate $\eta =\max (\eta _{\left\vert
00\right\rangle },\eta _{\left\vert 01\right\rangle },\eta _{\left\vert
10\right\rangle },\eta _{\left\vert 11\right\rangle })$ is calculated. In
the calculations, up to 3-photon processes are included. The $N$-photon
processes for $N>3$ have negligible effects. For the SQUID qubits with the
previously given DPs, the leakage is a function of WPs $x_{e1}$, $x_{e2}$,
and $\kappa $. In Fig. 2(a) we plot the leakage of the CNOT gate versus $%
x_{e2}$ and $\kappa $ for $x_{e1}=0.499$ and $x_{m0}=2\times 10^{-4}$. It is
shown that the leakage of the CNOT gate is much smaller when the coupled
SQUID qubits are operated around $x_{e2}=0.4997$ and $\kappa =7.5\times
10^{-4}$ for $x_{e1}=0.499$.

\begin{figure}[ptb]
\begin{center}
\psfig{file=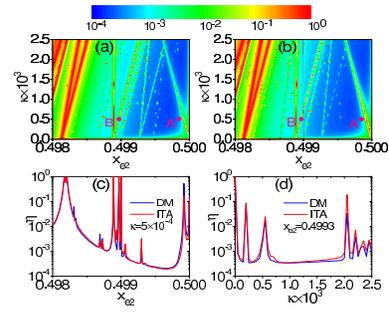,height=4cm,width=5cm}
\end{center}
\caption{Leakage of the CNOT gate for $x_{e1}=0.499$ and $x_{m0}=2\times
10^{-4}$. (a) the leakage vs. $x_{e2}$ and $\protect\kappa $ obtained from
ITA, (b) the leakage vs. $x_{e2}$ and $\protect\kappa $ obtained from DM,
(c) comparison of the leakage vs. $x_{e2}$ at $\protect\kappa =5\times
10^{-4}$, and (d) comparison of the leakage vs. $\protect\kappa $ at $%
x_{e2}=0.4993$.}
\end{figure}

To evaluate the results obtained from ITA, we performed an accurate dynamic
calculation by solving the TDSE of the coupled qubits $i\partial c_{n}\left(
\tau \right) /\partial \tau =\sum_{n^{\prime }}H_{nn^{\prime }}^{R}\left(
\tau \right) c_{n^{\prime }}\left( \tau \right) $ for the probability
amplitudes $c_{n}$ of the first 20 eigenstates $|n)$ using the
split-operator method \cite{Hermann1988}, where $\tau =\omega _{LC}t$ and $%
H_{nn^{\prime }}^{R}=\left[ E_{n}\delta _{nn^{\prime }}+\left( n\left\vert
V\right\vert n^{\prime }\right) \right] /\hbar \omega _{LC}$ \cite%
{zhou-asc2004}. The probability of being in the state $|n)$ is $\left\vert
c_{n}\right\vert ^{2}$, from which\ the maximum probabilities on the
undesired states and the leakage $\eta \left( x_{e1},x_{e2},\kappa \right) $
of the CNOT gate are calculated. The results are shown in Fig. 2(b) with the
color scale identical to Fig. 2(a). To have a more quantitative comparison,
we also plot, in Fig. 2(c) and (d), two line-cut figures at places with the
richest structures in Fig. 2(a) and (b). The good agreement between the
results of ITA and DM demonstrates the validity of using ITA to minimize the
CNOT gate leakage.

One of the advantages of ITA is that it provides clear physical intuition
and insight into the origin of intrinsic gate errors and thus how to reduce
the problem by selecting appropriate WPs. Furthermore, ITA is orders of
magnitude faster than DM. For instance, the result shown in Fig. 2(a) took
less than 3 hours to compute on a dual-2.8 GHz processor Dell PRECISION 650
workstation while that shown in Fig. 2(b) took more than 300 hours on the
same computer. Denoting $\tau _{S}$ and $\tau _{T}$ respectively the time
needed to calculate the spectroscopic properties of an $n$-bit gate and that
used to compute the probability evolution in DM for each component. The
total times needed to compute the leakage by ITA and DM are approximately $%
\tau _{I}\approx \tau _{S}$ and $\tau _{D}\approx \tau _{S}+2^{n}\tau _{T}$,
respectively, where the factor $2^{n}$ is the number of components for the $%
n $-bit gate. Hence, $\tau _{S}<\tau _{T}$ and the ratio $\tau _{D}/\tau
_{I}\approx 2^{n}\zeta $ increases exponentially with $n$, where $\zeta
\equiv \tau _{T}/\tau _{S}$. For weak fields, $\zeta \gg 1$ and the ratio $%
\tau _{D}/\tau _{I}$ is very large. For example, for the CNOT gate above $%
\zeta \sim 25$ and $\tau _{D}/\tau _{I}\sim 100$. Thus as the number of
qubits increases the optimization of WPs of multi-bit quantum gates could
become an intractable problem using DM.

\begin{figure}[ptb]
\begin{center}
\psfig{file=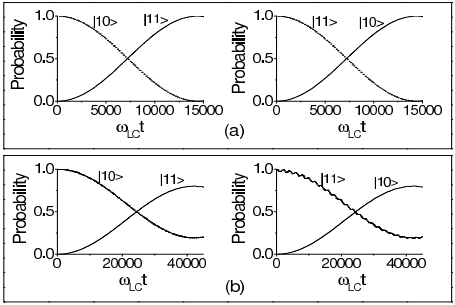,height=4cm,width=6cm}
\end{center}
\caption{Population evolution of the computational states $\left\vert
10\right\rangle $ and $\left\vert 11\right\rangle $ for the CNOT gates with
different WPs:\ (a) $x_{e2}=0.49985 $ and $\protect\kappa =5\times 10^{-4}$
corresponding to the point "A" in Fig. 2(a) [also (b)], and (b) $%
x_{e2}=0.49897$ and $\protect\kappa =5\times 10^{-4}$ corresponding to the
point "B" in Fig. 2(a) [also (b)]. In both cases, $x_{m0}=2\times 10^{-4}$
and $x_{e1}=0.499$.}
\end{figure}

To demonstrate quantitatively the effect of different WPs on intrinsic gate
errors, we plot in Fig. 3(a) and (b) the population evolution of the
computational states $\left\vert 10\right\rangle $ and $\left\vert
11\right\rangle $ for the CNOT gate operated with two different sets of WPs
corresponding to points "A" and "B" marked in Fig. 2(a) [also (b)],
respectively. It is clearly shown that inversion from the initial state $%
\left\vert 10\right\rangle $ ($\left\vert 11\right\rangle $) to the final
state $\left\vert 11\right\rangle $ ($\left\vert 10\right\rangle $) after
the $\pi $-pulse is almost complete when operated at the point "A" but
significantly incomplete at the point "B". We also computed the population
evolution of the coupled qubits with the initial states $\left\vert
00\right\rangle $ and $\left\vert 01\right\rangle $ using the same microwave
pulse. The system essentially stayed in the initial states for the gate
operated at the point "A" but leaked significantly to the non-computational
states for the gate at the point "B". The quality of a gate can be described
by gate fidelity $F\equiv \overline{\text{Trace}\left[ \rho _{P}\rho _{I}%
\right] }$, where $\rho _{P}$ and $\rho _{I}$ are the physical and ideal
density matrices after gate operation and the overline denotes averaging
over all possible initial states \cite{Li2003}. For the CNOT gates operated
at the point "A" we obtain $F_{A}=0.9997$ which is very close to the ideal
CNOT gate. In \ contrast, the gate operated at the point "B" has $%
F_{B}=0.8061$ which is too large to be tolerated. Furthermore, at the point
"A" the gate is about a factor of three faster than that at the point "B".

In summary, a very efficient method is proposed to optimize the WPs of
microwave-driven quantum logical gates in multi-level qubits based on ITA.
In this method, the leakage is estimated accurately from the spectroscopic
properties of the qubits and minimized by choosing appropriate WPs. This
method is exemplified by optimizing the WPs of coupled rf SQUID flux qubits
for minimal leakage of the CNOT gate. The result is in good agreement with
that obtained from dynamic calculations. Compared to the conventional
dynamic method the ITA not only provides physical insight into the origin of
gate leakage but also reduces the computational time by more than two orders
of magnitude for the CNOT gate. Furthermore, since the ratio of the speed of
ITA to that of DM scales exponentially as $2^{n}$ the ITA is scalable as the
number of qubit $n$ increases. Our calculation also shows that high
intrinsic fidelity CNOT gate can be achieved using microwave-driven rf SQUID
qubits with properly selected WPs. Although the proposed ITA is only
applied, as an example, to the rf SQUID qubits, it is also valid for other
microwave-driven multi-level qubits. Therefore, ITA provides a much needed
solution for the optimization of WPs of multi-bit gates implemented with
multi-level physical qubits, for which DM is extremely time consuming or
could even become intractable.

This work was supported in part by the NSF (DMR-0325551) and AFOSR, NSA and
ARDA through DURINT grant (F49620-01-1-0439).

\bibliographystyle{apsrev}
\bibliography{t-squid}

\end{document}